\documentclass[aps,prl,10pt,twocolumn,superscriptaddress,preprintnumbers]{revtex4}

\usepackage[dvipsnames]{xcolor}
\usepackage{physics}
\usepackage{bm}
\usepackage{braket}

\usepackage{scalerel}
\usepackage{blindtext}
\usepackage{epsfig, cancel}

\usepackage{latexsym}
\usepackage{natbib, comment}
\usepackage{mathrsfs,amsmath,amssymb,amsthm,amsfonts,tikz,graphicx,accents,hyperref,color, bbm}
\usepackage{url}
\usepackage{dcolumn}
\usepackage{multirow}
\usepackage{color}
\usepackage{cancel}
\usepackage{soul}
\usepackage[normalem]{ulem}
\usepackage{txfonts}
\usepackage{epsfig}
\usepackage{psfrag}
\usepackage{subfigure}
\hypersetup{colorlinks=true}
\usepackage{mathtools}
\usepackage{enumitem}
\usepackage{float}
\usepackage{caption,ragged2e}

\usepackage[all]{xy}
\usepackage{slashed}
\usepackage{slashed,ccaption}
\usepackage{multirow}

\usepackage{caption}

\usepackage{array}
\hypersetup{ linktoc=all,
    colorlinks, linkcolor={palatinateblue},
    citecolor={red}, urlcolor={amaranth} 
}

\graphicspath{{Images/}}

\DeclareSymbolFont{extraup}{U}{zavm}{m}{n}
\DeclareMathSymbol{\varheart}{\mathalpha}{extraup}{86}
\DeclareMathSymbol{\vardiamond}{\mathalpha}{extraup}{87}
\makeatletter
\renewcommand*{\@fnsymbol}[1]{\ensuremath{\ifcase#1\or \clubsuit \or \vardiamond \or \varheart\or
    \spadesuit\or \mathparagraph\or \|\or **\or \dagger\dagger
    \or \ddagger\ddagger \else\@ctrerr\fi}}
\makeatother

\definecolor{rosy}{RGB}{230,235,252}
\definecolor{myframetitle}{RGB}{90,89,170}
\definecolor{myblocktitle}{RGB}{140,185,249}
\definecolor{mytitle}{RGB}{10,80,26}

\definecolor{darkgreen}{RGB}{27,130,45}
\definecolor{darkblue}{rgb}{0,0,0.3}
\definecolor{darkred}{rgb}{0.7,0,0}

\definecolor{light gray}{RGB}{220,220,220}
\definecolor{dark purple}{RGB}{108,0,217}
\definecolor{pink}{RGB}{190,20,100}
\definecolor{orang}{RGB}{193,63,0}
\definecolor{green}{RGB}{11,98,17}
\definecolor{darkpink}{RGB}{153,0,76}
\definecolor{bluegreen}{RGB}{0,102,102}
\definecolor{greenlagan}{RGB}{0,102,0}
\definecolor{redgreen}{RGB}{102,102,0}
\definecolor{Redgreen}{RGB}{153,76,0}
\definecolor{vividviolet}{rgb}{0.62, 0.0, 1.0}
\definecolor{amaranth}{rgb}{0.9, 0.17, 0.31}
\definecolor{palatinateblue}{rgb}{0.15, 0.23, 0.89}
\definecolor{brightpink}{rgb}{1.0, 0.0, 0.5}
\definecolor{cornflowerblue}{rgb}{0.39, 0.58, 0.93}
\definecolor{deepcarminepink}{rgb}{0.94, 0.19, 0.22}
\definecolor{radicalred}{rgb}{1.0, 0.21, 0.37}

\usepackage{centernot}
\usepackage{tikz}
\usetikzlibrary{decorations.pathmorphing,backgrounds,shapes,arrows,shadows}
\tikzset{zigzag/.style={decorate,decoration=zigzag}}
\tikzset{snake it/.style={decorate, decoration=snake}}

\newcommand{\bec}{\begin{center}}
\newcommand{\eec}{\end{center}}

\def\L{\mathcal{L}}

\def\bea#1\ena{\begin{align}#1\end{align}}

\def\a{\alpha}
\def\ta{\tilde\alpha}
\def\b{\beta}
\def\tb{\tilde\beta}

\def\checkmark{\tikz\fill[scale=0.4](0,.35) -- (.25,0) -- (1,.7) -- (.25,.15) -- cycle;}

\newcommand\tcb{\textcolor{blue}}

\hypersetup{ linktoc=all,
    colorlinks=true, linkcolor={red},
    citecolor={blue}, urlcolor={magenta}
}

\def\dett[#1,#2,#3,#4]{\left|%
\begin{array}{cc} #1 & #2 \\ #3 & #4 \end{array} \right|}
\def\matt[#1,#2,#3,#4]{\left(%
\begin{array}{cc} #1 & #2 \\ #3 & #4 \end{array} \right)}
\def\mattt[#1,#2,#3,#4,#5,#6,#7,#8,#9]{\left(%
\begin{array}{ccc} #1 & #2 & #3 \\ #4 & #5 & #6 \\ #7 & #8 & #9 \end{array} \right)}
\def\dettt[#1,#2,#3,#4,#5,#6,#7,#8,#9]{\left|%
\begin{array}{ccc} #1 & #2 & #3 \\ #4 & #5 & #6 \\ #7 & #8 & #9 \end{array} \right|}
\def\vect[#1,#2]{\left(%
\begin{array}{cc} #1 \\ #2  \end{array} \right)}
\def\vectt[#1,#2]{\left(%
\begin{array}{cc} #1 \\ #2  \end{array} \right)}
\def\vecttt[#1,#2,#3]{\left(%
\begin{array}{cc} #1 \\ #2 \\ #3 \end{array} \right)}
\def\sqvect[#1,#2]{\left[%
\begin{array}{cc} #1 \\ #2  \end{array} \right]}
\def\wbvect[#1,#2]{\left\{%
\begin{array}{cc} #1 \\ #2  \end{array} \right\}}
\def\tvect[#1,#2]{\left(%
\begin{array}{cc} #1 & #2  \end{array} \right)}
\def\tvectt[#1,#2]{\left(%
\begin{array}{cc} #1 & #2  \end{array} \right)}
\def\tvecttt[#1,#2,#3]{\left(%
\begin{array}{ccc} #1 & #2 & #3 \end{array} \right)}
\def\ket[#1]{\left| #1 \right\rangle}
\def\bra[#1]{\left\langle #1 \right|}
\def\brak[#1,#2]{\left\langle #1 | #2 \right\rangle}
\def\braket[#1,#2]{\left\langle#1\right.\hspace{-2.pt}\left| #2\right\rangle}
\def\pair[#1,#2]{\left\langle #1 , #2 \right\rangle}
\def\be{\begin{equation}}
\def\ee{\end{equation}}
\def\ben{\begin{equation*}}
\def\een{\end{equation*}}
\usepackage{mathtools}

\newcommand{\DeclareAutoPairedDelimiter}[3]{%
	\expandafter\DeclarePairedDelimiter\csname Auto\string#1\endcsname{#2}{#3}%
	\begingroup\edef\x{\endgroup
		\noexpand\DeclareRobustCommand{\noexpand#1}{%
			\expandafter\noexpand\csname Auto\string#1\endcsname*}}%
	\x}
\DeclareAutoPairedDelimiter{\p}{(}{)}

\usepackage[most]{tcolorbox}

\tcbset{highlight math style={left=02mm,right=02mm,top=-1mm,bottom=-1mm}} 
\usepackage{empheq}

\begin{document}

\title{Strings, Virasoro Sandwiches and Worldsheet Horizons}

\author{Arjun Bagchi}\email{abagchi@iitk.ac.in}\affiliation{Indian Institute of Technology Kanpur, Kalyanpur, Kanpur 208016, INDIA}
\author{Aritra Banerjee}\email{aritra.banerjee@pilani.bits-pilani.ac.in}\affiliation{Birla Institute of Technology and Science, Pilani Campus, Pilani, Rajasthan 333031, INDIA }
\author{Ida M. Rasulian}\email{idarasulian@ipm.ir}
\affiliation{ School of Physics, Institute for Research in Fundamental
Sciences (IPM), P.O.Box 19395-5531, Tehran, IRAN}
\author{M. M. Sheikh-Jabbari}\email{jabbari@theory.ipm.ac.ir}
\affiliation{ School of Physics, Institute for Research in Fundamental
Sciences (IPM), P.O.Box 19395-5531, Tehran, IRAN}

\begin{abstract}
{We revisit the canonical quantization of free bosonic closed string theory and observe that the physicality of states requires vanishing of the worldsheet Virasoro algebra generators sandwiched between any two physical states. This requirement yields four classes of physical states, depending on discrete worldsheet symmetries: parity and time reversal. The usual string states which are highest weight states of the Virasoro algebra, preserve both, while the other new three classes  break one or both. We apply our formulation to an accelerated worldsheet with horizons, initiating the worldsheet formulation of a thermal string theory and strings probing horizon of black holes.}

 \end{abstract}

\maketitle

A complete quantum theory of gravity still eludes us. Of the vying candidates of quantum gravity, string theory is arguably the leading one 
\cite{Green:1987sp, Polchinski:1998rq}. Using special relativity and quantum mechanics as inputs in a theory of fundamental one-dimensional extended objects, string theory naturally generates a theory of quantum gravity.  The ``old covariant quantization''  is one of the most useful methods used to quantize strings \cite{Green:1987sp}. As we  review below, this uses a partial gauge fixing on the worldsheet of the string to bring the worldsheet theory to a 2d conformal field theory where highest weight representations of the Virasoro algebra play a central role. In this letter, we carefully revisit this quantization of the theory in the context of closed strings and find surprising new results. 

Our results hinge on the important observation that any classical constraint $\mathcal{O}=0$  should be imposed as $\langle \psi'| \mathcal{O}|\psi\rangle =0$ for any $|\psi\rangle, |\psi'\rangle$ in the physical Hilbert space of the theory. We call this the ``sandwich condition''. In the case of string theory,  this general statement should be applied  to the Virasoro generators which are the modes of the worldsheet stress tensor. This generalises earlier textbook implementation of the constraints where this was restricted to the Virasoro highest weight conditions \cite{Green:1987sp, Polchinski:1998rq}. We show that consistent implementation of the Virasoro sandwich condition leads to four different classes of states, depending on whether or not discrete symmetries on the worldsheet, viz. parity or time-reversal are preserved. The textbook formulation of string theory only considers what we will call \textbf{Class I} states that preserve both discrete symmetries. The other three classes break one (\textbf{Class II} and \textbf{III}) or both (\textbf{Class IV}). 

We go on to exemplify our construction by considering non-inertial string worldsheets. These worldsheets have an intrinsic acceleration associated with them and hence Rindler horizons form on the worldsheet. Non-inertial worldsheets are of importance when one considers strings at finite temperature or strings near black holes. We show that in order to describe physics on these worldsheets, we need to have states beyond conventional \textbf{Class I} states. 

{We emphasise here that this is merely one of the possible applications of our rather general framework, which opens the door to the formulation of thermal string theory in a manner parallel to thermal quantum field theory from the perspective of the worldsheet. }

\bec
\textbf{Quick Review of Textbook  Worldsheet Theory}\label{sec:review}
\eec
String theory on a $D$ dimensional target space is defined in terms of the Polyakov action \cite{Polchinski:1998rq}
\be\label{worldsheet-action}
\text{S}_P= \int d^2\sigma \sqrt{\gamma}\gamma^{ab} \partial_a X^\mu \partial_b X^\nu G_{\mu\nu} 
\ee
where we have put the string tension $T=1/2\pi\alpha' =1$. 
This action enjoys Diff\! $\oplus$\! Weyl gauge symmetry: any two worldsheet metrics that are equivalent up to Weyl scaling or 2d diffeomorphisms, are physically equivalent at classical level.  
One can use these symmetries to \textit{locally} fix any worldsheet metric  to Minkowski metric,
\be\label{gamma-eta}
\gamma^{ab} = \eta^{ab},
\ee
i.e, worldsheet geometry may be set to Minkowski up to a chosen topology. 
Equation of motion in the conformal gauge in a flat target space is simply the wave equation in worldsheet coordinates $\tau, \sigma$ .

The solutions for a closed string with $\sigma\sim \sigma+2\pi$ periodicity can be mode expanded in terms of left-moving and right-moving waves
\be\label{string-mode-expansion-alpha}
X^\mu = x^\mu +  \tilde{x}^\mu + \alpha_0^\mu \sigma_++ \ta_0^\mu\sigma_-+ \sum_{n\neq 0} \frac{1}{n}\left(\a^\mu_n e^{in\sigma_+} + \ta^\mu_ne^{in\sigma_-} \right),
\ee
where $\sigma_\pm = \tau \pm \sigma$.
To avoid cluttering, we will suppress the target space $\mu$ superscript and retain it only when crucial. Consistency of  fixing the gauge \eqref{gamma-eta} requires that the worldsheet energy-momentum tensor should also vanish on-shell:
\be\label{classical-constraint}
T_{ab}:=-\frac{2}{\sqrt{\gamma}} \frac{\delta \text{S}_P}{\delta \gamma^{ab}}=0.
\ee
One can perform canonical quantization by promoting $\a_n, \ta_n$ to operators satisfying commutation relations
\be\label{inertial-osci}
\begin{split}[\a_n, \a_m]= {\frac{n}{2}} \delta_{n+m,0}=[\ta_n, \ta_m]
,\qquad [\a_n, \ta_m]= 0,\\
[x, \a_n]=\frac{i}{2}\delta_{n,0}=[\tilde{x},\ta_n]
,\qquad [\tilde{x}, \alpha_n]=[x,\ta_n]=0, 
\end{split}\ee
 and choosing an appropriate vacuum state with center of mass momentum $k^\mu$, 
\be\label{avac}
\begin{split}
\a^\mu_n |0; k^\mu \rangle_{_\a} = 0 = \ta^\mu_n |0; k^\mu \rangle_{_\a}, \quad \forall n>0,\\
\a^\mu_0 |0; k^\mu \rangle_{_\a}=\ta^\mu_0 |0; k^\mu \rangle_{_\a}=\frac12 k^\mu |0; k^\mu \rangle_{_\a}.
\end{split}\ee
The most general state is then constructed by the action of generic combinations of powers of $\a_n, \ta_n$ with $n<0$ on the vacuum. Denote the Hilbert space of all such states by ${\cal H}_{\text{\tiny{T}}}$. Physical states are the subset ${\cal H}_{\text{\tiny{P}}}$ of ${\cal H}_{\text{\tiny{T}}}$ that satisfy \cite{Polchinski:1998rq}
\be\label{Sandwich-constriants-Tab}
\langle phys' |T_{ab}|phys\rangle=0 \qquad \forall |phys\rangle, |phys'\rangle\in {\cal H}_{\text{\tiny{P}}}\ .
\ee
Nonzero components of $T_{ab}$ in gauge \eqref{gamma-eta} are
\be\label{L0L0b-alpha}
\begin{split}
T_{++} = \sum_n \L^+_{n} e^{in\sigma_+} &,\ \L^+_n = \sum_{m=-\infty}^\infty \a_{n-m}\cdot \a_m - a \ \delta_{n,0}\\ 
\quad T_{--} = \sum_n \L^-_{n} e^{in\sigma_-} &, \ \L^-_n = \sum_{m=-\infty}^\infty \ta_{n-m}\cdot \ta_m- a \ \delta_{n,0}.
\end{split}\ee
$\L^\pm_n$ generate two copies of the Virasoro algebra and  $a=\frac{D-2}{12}$ is the normal ordering constant, the zero point energy. Note that in our conventions $\L^\pm_0$ denotes what in a more usual conventions is $\L^\pm_0 + a$. The residual symmetries of the worldsheet symmetry after gauge fixing \eqref{gamma-eta} is the 2d conformal algebra, 
\be
[\L^\pm_n, \L^\pm_m] = (n-m) \L^\pm_{n+m} + \frac{c}{12} \delta_{n+m,0} (n^3-n), \ c=D-2
\ee 
with $[\L_n^\pm, \L_m^\mp]=0$. 

\bec
\textbf{Sandwich Virasoro  Constraints}\label{sec:SVC}
\eec

Constraints \eqref{Sandwich-constriants-Tab}  yield the ``Sandwich Virasoro Conditions'' (SVC), for all states $|phys\rangle, | phys'\rangle\in {\cal H}_{\text{\tiny{P}}}$
\be\label{SVC}
{\langle phys| \L_n^+ | phys'\rangle = 0 = \langle phys| \L_n^- | phys'\rangle \qquad \forall n\in \mathbb{Z}}.
\ee
In almost all string theory textbooks, e.g. \cite{Green:1987sp, Polchinski:1998rq}, the above condition has been solved through the right-action, 
\be\label{ra}
\L^\pm_n | phys\rangle = 0, \quad \forall n \geq 0. 
\ee 
{All solutions to \eqref{ra} also automatically satisfy  \eqref{SVC}. But, as we emphasize below, all solutions of \eqref{SVC} don't necessarily obey \eqref{ra}\footnote{Related developments arose out of covariant quantization of the null (or tensionless) string worldsheet, where such sandwiched conditions are necessary for consistency \cite{Bagchi:2020fpr}. See also \cite{Banerjee:2023ekd, Banerjee:2024fbi}.}. We now focus on these other solutions of \eqref{SVC}. By construction for all physical states $|\psi \rangle, |\psi' \rangle$, we would have,
\be
\L^\pm_n |\psi \rangle \neq 0, \quad \text{but} \quad \langle \psi'|\L^\pm_n |\psi \rangle =0.
\ee
{All states $|\phi_i\rangle$ that satisfy $\L_0^\pm|\phi_i\rangle=0$, automatically satisfy $\langle\phi_i|\L_n^\pm|\phi_j\rangle=0$, by virtue of the Virasoro algebra and hermiticity of $\L^\pm_0$ \footnote{To see this explicitly, recall that $\L_n=\frac1n[\L_n,\L_0]$ and that $\langle\phi_i|[\L_n,\L_0]|\phi_j\rangle=0$ as $\L_0|\phi_i\rangle=0$ and $\langle\phi_i|\L_0=0$.}. So, in our quest to find the solutions to SVC which are non-highest weights of the Virasoro algebra, we can thus focus on only the $n=0$ sector: 
\be\label{L0BL0-Sand--Const}
\L^\pm_0 |\psi \rangle \neq 0, \quad \text{but} \quad \langle \psi'|\L^\pm_0 |\psi \rangle =0.
\ee
Also note that all physical states are defined up to spurious states, which are defined as all states $|\chi\rangle$ with $\L_0^\pm|\chi\rangle=0$ and $\langle\psi|\chi\rangle=0$, for any $|\psi\rangle$, s.t. $\L_n^\pm|\psi\rangle=0, \forall n\geq 0$. Our discussions above imply that, as in the right-action case \eqref{ra},  focusing on \eqref{L0BL0-Sand--Const} specifies physical states up to spurious states which should be dealt with, as in usual string theory \cite{Green:1987sp, Polchinski:1998rq}.  }

\textit{Classification of solutions to SVC.} {We will prefer to work with combinations $\L_0^{+}\pm {\L}_0^{-}$ over $\L^\pm_0$.} Recall that $\L_0^{+}\pm {\L}_0^{-}$ are well-defined self-adjoint (Hermitian) operators over  ${\cal H}_{\text{\tiny{T}}}$. However, not all states in ${\cal H}_{\text{\tiny{T}}}$ are physical; one can divide ${\cal H}_{\text{\tiny{T}}}$ into the physical Hilbert space ${\cal H}_{\text{\tiny{P}}}$ and its complement ${\cal H}_{\text{\tiny{C}}}$, 
\begin{equation}\label{HpHc}
\hspace*{-3mm}{\cal H}_{\text{\tiny{T}}}= {\cal H}_{\text{\tiny{P}}} \cup {\cal H}_{\text{\tiny{C}}},\  \text{s.t.} \ \langle \psi_{\text{\tiny{P}}}|\psi_{\text{\tiny{C}}}\rangle=0,\ \forall |\psi_{\text{\tiny{P}}}\rangle\in {\cal H}_{\text{\tiny{P}}}, \ |\psi_{\text{\tiny{C}}}\rangle\in {\cal H}_{\text{\tiny{C}}}. 
\end{equation} 
This split of the Hilbert space into two is reminiscent of what naturally happens in spacetimes which have a horizon, e.g. a black hole spacetime where a non-inertial  observer has access to only states that lie outside the event horizon. Later in the paper, when we focus on a class of examples dealing with a non-inertial string, we will see that our construction can be interpreted in terms of appearance of a horizon, this time on the string worldsheet itself.

We now get back to the job at hand, finding the solutions to SVC. We find the most general solutions to \eqref{L0BL0-Sand--Const} fall into either of the  following \textit{four} classes:
\begin{enumerate}
\item[\textbf{I}.] $(\L_0^+-\L_0^{-}) |phys\rangle =0,\qquad (\L_0^{+}+\L_0^{-}) |phys\rangle =0$; 
\item[\textbf{II}.] $(\L_0^+-\L_0^{-}) |phys\rangle \in {\cal H}_{\text{\tiny{C}}},\qquad (\L_0^{+}+\L_0^{-}) |phys\rangle =0$;

\item[\textbf{III}.] $(\L_0^+-\L_0^{-}) |phys\rangle =0,\qquad (\L_0^{+}+\L_0^{-}) |phys\rangle \in {\cal H}_{\text{\tiny{C}}}$; 
\item[\textbf{IV}.] $(\L_0^+-\L_0^{-})|phys\rangle \in {\cal H}_{\text{\tiny{C}}},\qquad (\L_0^{+}+\L_0^{-}) |phys\rangle \in {\cal H}_{\text{\tiny{C}}}$; 
\end{enumerate}
Solutions of \eqref{ra} are all covered in \textbf{Class I}, while \textbf{Class II, III} and \textbf{IV} do not have any counterparts in the usual closed strings Hilbert space.

\textit{Construction of general solutions.} To explicitly construct ${\cal H}_{\text{\tiny{P}}}$ and \textbf{Class I--IV} cases therein, recall that $(\L_0^{+}\pm\L_0^{-})$ are two mutually commuting Hermitian operators on ${\cal H}_{\text{\tiny{T}}}$. Thus, their eigenstates provide a complete basis for ${\cal H}_{_{\text{T}}}$. Assuming their spectrum span over positive, zero and negative eigenvalues,
\be\label{L0bL0-eigenstates}
(\L_0^{+}\pm\L_0^{-}) |\lambda_+, \mathfrak{s}_+; \lambda_-, \mathfrak{s}_- \rangle= \mathfrak{s}_\pm\lambda_\pm |\lambda_+, \mathfrak{s}_+; \lambda_-, \mathfrak{s}_- \rangle,\ \  \lambda_\pm \geq 0
\ee
where $\lambda_{\pm}$ are the two eigenvalues of these operators and $\mathfrak{s}_\pm$ are two signs (equal to $\pm 1$). 
As already mentioned, \textbf{Class I} corresponds to $\lambda_\pm = 0$, \textbf{Class II} to $\lambda_+=0,\lambda_-\neq 0$,  \textbf{Class III} to $\lambda_-=0,\lambda_+\neq 0$ and finally \textbf{Class IV} to both $\lambda_\pm\neq 0$. Now on the other hand, to solve for the $(\L_0^{+}\pm\L_0^{-}) |phys\rangle \in {\cal H}_{\text{\tiny{C}}}$ conditions, consider a full set of states of the form
\be\label{s-12-states}
\begin{split}
|\lambda_\pm ;s_1,s_2\rangle &:= {\cal N}\bigg[|\lambda_+, +; \lambda_-, + \rangle + s_1 |\lambda_+, -; \lambda_-, + \rangle \\ &+s_2 |\lambda_+, +; \lambda_-, - \rangle+ s_1 s_2 |\lambda_+, -; \lambda_-, - \rangle\bigg],
\end{split}
\ee
where $s_1,s_2$ may be chosen to be $0, \pm 1$ and 
\be\label{s-12-states-L0bL0}
\begin{split}
(\L_0^{+}-\L_0^{-})|\lambda_\pm ;s_1,s_2\rangle &=\lambda_-\ |\lambda_\pm ;s_1,-s_2\rangle,\\ (\L_0^{+}+\L_0^{-})|\lambda_\pm ;s_1,s_2\rangle &=\lambda_+\ |\lambda_\pm ;-s_1,s_2\rangle,
\end{split}
\ee
with  ${\cal N}$ chosen such that,
\be
\langle \lambda'_\pm ;s'_1,s'_2|\lambda_\pm ;s_1,s_2\rangle= \delta_{\lambda,\lambda'}\delta_{s_1s'_1}\delta_{s_2s'_2}.
\ee
For independent set of states we can restrict to cases with $s_1, s_2$ to be 0 or 1, for which  the constraints \eqref{L0BL0-Sand--Const} are already satisfied. Explicitly, 
\begin{subequations}\label{Class-I-II-III-IV-explicit}
\begin{align}
&\textbf{Class I}
: \qquad |0,+;0, + \rangle \label{Class-I-explicit}\\
&\textbf{Class II}:
\quad   \frac{1}{\sqrt{2}} \left(|0, +; \lambda_- ;+\rangle+ |0, +; \lambda_- ;-\rangle \right)\label{Class-II-explicit}\\
&\textbf{Class III}: 
~~  \frac{1}{\sqrt{2}} \left(|\lambda_+, +; 0 ;+\rangle+ |\lambda_+, -; 0 ;+\rangle \right) \label{Class-III-explicit}\\
&\textbf{Class IV}:
\ \ \ \frac12\bigg[|\lambda_+, +; \lambda_-, + \rangle + |\lambda_+, -; \lambda_-, + \rangle \nonumber\\ &\hspace*{20mm} + |\lambda_+, +; \lambda_-, - \rangle+  |\lambda_+, -; \lambda_-, - \rangle\bigg].\label{Class-IV-explicit}
\end{align}
\end{subequations}
Here $s_1,s_2$ can be associated with a $\mathbb{Z}_2\times \mathbb{Z}_2$ symmetry that divides the total Hilbert space ${\cal H}_{_{\text{T}}}$ into four sectors, as summarized in the Table \ref{Z2Z2-table}.

\begin{table}[h!]
\centering
 \begin{tabular}{||c c c c||} 
 \hline
 States & ~~($s_1,s_2$)~~ & $\mathbb{Z}_2 (\text{P})$  & $\mathbb{Z}_2 (\text{T})$ \\ [0.5ex] 
 \hline\hline
 \textbf{Class I} & (0,0) & \checkmark & \checkmark \\ 
 \hline\hline
 \textbf{Class II} & (1,0) & $\cross$ & \checkmark  \\
 \hline\hline
 \textbf{Class III} & (0,1) & \checkmark & $\cross$  \\
 \hline\hline
 \textbf{Class IV}& (1,1) & $\cross$ & $\cross$ \\ [0.5ex] 
 \hline
 \end{tabular}
 \caption{A summary of the characteristics for four classes of SVC solutions.}\label{Z2Z2-table}
\end{table}
Put differently, constraints \eqref{ra} are equivalent to SVC \eqref{SVC} supplemented with the full $\mathbb{Z}_2\times \mathbb{Z}_2$ symmetry. We show in a upcoming section that this $\mathbb{Z}_2\times \mathbb{Z}_2$ symmetry has a precise geometric meaning on a Rindler space, i.e. on a non-inertial worldsheet, where the two $\mathbb{Z}_2$'s would be associated with worldsheet parity P and time-reversal T.

We close this part by a further comment on the $\mathbb{Z}_2\times \mathbb{Z}_2$ structure. From \eqref{s-12-states-L0bL0} one learns that 
\be\label{Lpm-2}
(\L_0^{+}\pm\L_0^{-})^2 |\lambda_\pm ;s_1,s_2\rangle=\lambda^2_\pm\ |\lambda_\pm ;s_1,s_2\rangle.
\ee
That is $|\lambda_\pm ;s_1,s_2\rangle$, while not eigenstates of $\L_0^{+}\pm\L_0^{-}$, are eigenstates of $(\L_0^{+}\pm\L_0^{-})^2$. In a different wording, for either of the four Classes, $(\L_0^{+}\pm\L_0^{-})^2|\psi^p\rangle\in {\cal H}_{\text{\tiny{P}}}, (\L_0^{+}\pm\L_0^{-})^2|\psi^c\rangle\in {\cal H}_{\text{\tiny{C}}}$. 

\bec
\textbf{Example of  Quantized String Theory}\label{sec:Example}
\eec

So far the arguments were not specific to the string theory Hilbert space constructed through \eqref{avac}, for which we recall:
\be\label{L0L0b-string}
\begin{split}
\L_0^{+}|r,s;k^\mu\rangle = (r-\frac{m^2}{4} -a)|r,s; k^\mu\rangle,\\ \L_0^{-}|r,s; k^\mu\rangle = (s-\frac{m^2}{4}-a)|r,s;k^\mu\rangle, 
\end{split}\ee
where $a$ is the zero point energy in \eqref{L0L0b-alpha},  $r,s$ are non-negative integers, $k^\mu$ denotes the momentum of center of mass of strings and $m^2:=-k^2$; $k^\mu$ is taken to be a causal vector, i.e.  $m^2\geq 0$ \footnote{There is a tachyon in the usual bosonic string spectrum that can be included by taking $m^2\geq -4$.}. Therefore,  both $\L_0^+ + \L_0^-$ and $\L_0^+-\L_0^-$ have  positive and negative  eigenvalues that for the former they can be any real number and for the latter  are integer-valued, fulfilling the assumption asserted above \eqref{L0bL0-eigenstates}. Explicitly, in the notation of the previous part, 
\be\label{lambda-pm-rs}
\lambda_-=|r-s|,\qquad \lambda_+=|r+s-\frac{m^2}{2}-2a|.
\ee

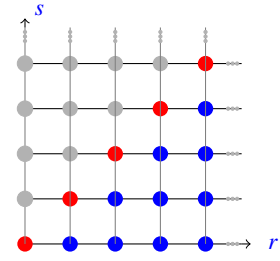
\begin{figure}[t]
\centering
\begin{tikzpicture}[scale=0.6]
\draw[black, thin] (0,0) -- (4.8,0);
\fill[red] (0,0) circle (0.17);
\fill[blue] (1, 0) circle (0.17);
\fill[blue] (2, 0) circle (0.17);
\fill[blue] (3, 0) circle (0.17);
\fill[blue] (4, 0) circle (0.17);
\filldraw[black!30] (4.5,0) circle (1pt);
\filldraw[black!30] (4.6,0) circle (1pt);
\filldraw[black!30] (4.7,0) circle (1pt);
\draw[->, black] (4.8,0) -- (5,0);
\draw[blue] (5.2,0) node[right] (script) {$r$};

\draw[black, thin] (0,1) -- (4.8,1);
\fill[red] (1,1) circle (0.17);
\fill[blue] (2,1) circle (0.17);
\fill[blue] (3,1) circle (0.17);
\fill[blue] (4,1) circle (0.17);
\filldraw[black!30] (4.5,1) circle (1pt);
\filldraw[black!30] (4.6,1) circle (1pt);
\filldraw[black!30] (4.7,1) circle (1pt);

\draw[black, thin] (0,2) -- (4.8,2);
\fill[black!30] (1,2) circle (0.17);
\fill[red] (2,2) circle (0.17);
\fill[blue] (3,2) circle (0.17);
\fill[blue] (4,2) circle (0.17);
\filldraw[black!30] (4.5,2) circle (1pt);
\filldraw[black!30] (4.6,2) circle (1pt);
\filldraw[black!30] (4.7,2) circle (1pt);

\draw[black, thin] (0,3) -- (4.8,3);
\fill[black!30] (1,3) circle (0.17);
\fill[black!30] (2,3) circle (0.17);
\fill[red] (3,3) circle (0.17);
\fill[blue] (4,3) circle (0.17);
\filldraw[black!30] (4.5,3) circle (1pt);
\filldraw[black!30] (4.6,3) circle (1pt);
\filldraw[black!30] (4.7,3) circle (1pt);

\draw[black, thin] (0,4) -- (4.8,4);
\fill[black!30] (1,4) circle (0.17);
\fill[black!30] (2,4) circle (0.17);
\fill[black!30] (3,4) circle (0.17);
\fill[red] (4,4) circle (0.17);
\filldraw[black!30] (4.5,4) circle (1pt);
\filldraw[black!30] (4.6,4) circle (1pt);
\filldraw[black!30] (4.7,4) circle (1pt);

\draw[gray, thin] (0,0) -- (0,4.8);
\filldraw[black!30] (0,1) circle (0.17);
\filldraw[black!30] (0,2) circle (0.17);
\filldraw[black!30] (0,3) circle (0.17);
\filldraw[black!30] (0,4) circle (0.17);
\filldraw[black!30] (0,4.5) circle (1pt);
\filldraw[black!30] (0,4.6) circle (1pt);
\filldraw[black!30] (0,4.7) circle (1pt);
\draw[->, black] (0,4.8) -- (0,5);
\draw[blue] (0,5.2) node[right] (script) {$s$};

\draw[gray, thin] (1,0) -- (1,4.8);
\filldraw[black!30] (1,4.5) circle (1pt);
\filldraw[black!30] (1,4.6) circle (1pt);
\filldraw[black!30] (1,4.7) circle (1pt);
\draw[gray, thin] (2,0) -- (2,4.8);
\filldraw[black!30] (2,4.5) circle (1pt);
\filldraw[black!30] (2,4.6) circle (1pt);
\filldraw[black!30] (2,4.7) circle (1pt);

\draw[gray, thin] (3,0) -- (3,4.8);
\filldraw[black!30] (3,4.5) circle (1pt);
\filldraw[black!30] (3,4.6) circle (1pt);
\filldraw[black!30] (3,4.7) circle (1pt);

\draw[gray, thin] (4,0) -- (4,4.8);
\filldraw[black!30] (4,4.5) circle (1pt);
\filldraw[black!30] (4,4.6) circle (1pt);
\filldraw[black!30] (4,4.7) circle (1pt);

\end{tikzpicture}
\caption{Physical string spectrum on a $\mathbb{Z}_2\times\mathbb{Z}_2$ lattice. The diagonal line (red dots) represents \textbf{Class I} states, for which we have $r=s$. The lower triangle (dark blue points) represent \textbf{Class II} or \textbf{III} states. \textbf{Class IV} states are not depicted here, as they are specified by 3 ordered non-negative  integers. }\label{Fig-1}
\end{figure}

\textbf{Class I} states have $\lambda_\pm=0$ \eqref{Class-I-explicit} are the usual level-matched string states of the form 
\begin{equation}\label{Class-I-strings}
|r,r;k^\mu\rangle,\qquad m^2=4(r-a)
\end{equation} 
and are hence labeled by a single non-negative integer (see Fig. \ref{Fig-1}). 

\textbf{Class II} states have $\lambda_+=0$ and $\lambda_-\neq 0$  \eqref{Class-II-explicit}:
\begin{equation}\label{Class-II-strings}
|r,s;k^\mu\rangle +  |s,r;k^\mu\rangle, \qquad m^2=2(r+s)-4a, \qquad r>s
\end{equation}
So they are labeled by two non-negative  integers $r>s$. 
\footnote{Note that, following the classification of SVC solutions, one can see the action of $(\L_0^+-\L_0^{-})$ transforms the above state to  $\sim |r,s;k^\mu\rangle -  |s,r;k^\mu\rangle$ which resides in the complement. }

\textbf{Class III} states have $\lambda_-=0$ and $\lambda_+\neq 0$, states of the form 
\begin{equation}\label{Class-III-strings}
|r,r;k^\mu\rangle + |s,s;k^\mu\rangle,\qquad   
m^2=2(r+s)-4a, \qquad r>s
\ee  
So they are labeled by two non-negative integers $r>s$. Note that both component states do not need to be stringy mass eigenstates on their own.

\textbf{Class IV} states have $\lambda_+, \lambda_-\neq 0$  \eqref{Class-IV-explicit} and are generically  quantified by three independent  ordered integers $r>s>|q|$:
\be\label{Class-IV-strings}
\begin{split}
&\frac{1}{2}(|r_1,s_1; k^\mu\rangle+ |r_2,s_2;k^\mu\rangle+|s_1,r_1;k^\mu\rangle+ |s_2,r_2;k^\mu\rangle),\\
&r_1=r+q,\ \ s_1=s+q,\ \ r_2=r-q,\ \ s_2=s-q,
\end{split}
\ee
or $2r=r_1+r_2,  2s=s_1+s_2, 2q=r_1-r_2=s_1-s_2$, with $m^2= 2(r+s)-4a$. Without loss of generality one can take $q>0$.

\bec
\textbf{$\mathbb{Z}_2\times \mathbb{Z}_2$ as Non-inertial Worldsheet Symmetry} 
\eec

{Our construction above may seem formal. To elucidate, we now focus on an explicit class of string worldsheets which will realize our above general construction. For this we will consider strings with intrinsic worldsheet acceleration, which are expected to occur when one considers strings that are close to black holes or when one heats up a gas of strings. These accelerated worldsheets have a natural Rindler structure induced on them and hence also form (observer dependent) horizons on the worldsheet \cite{Bagchi:2020ats, Bagchi:2021ban}. This situation is where  the split of Hilbert space mentioned in \eqref{HpHc} is realized.}

{We begin with classical considerations.} Fig.~\ref{fig:Quadrants} depicts the Penrose diagram of 2d Minkowski space that is a 4-fold cover of Rindler or Milne spaces and their metric are related as 
\begin{equation}
\begin{split}
\hspace*{-1cm}ds^2&= -dt^2+ dx^2,\quad \qquad\  \quad x,t\in \mathbb{R}, \qquad \text{Minkowski}
\\ 
 &= e^{\pm 2\kappa \sigma} (-d\tau^2 + d\sigma^2),\quad  \sigma, \tau \in \mathbb{R},  \quad \text{Right/Left Rindler},\\
      &= e^{\pm 2\kappa \tau} (-d\tau^2 + d\sigma^2),\quad  \sigma, \tau \in \mathbb{R},  \quad \text{Top/Down Milne},\nonumber
\end{split}
\end{equation}
where the Right/Left Rindler coordinates in terms of the original Minkowski  are given by:
\begin{equation}\label{x-t-sigma-tau}
x\pm t =\left\{\begin{array}{cc}\frac{1}{\kappa} \ e^{\kappa(\sigma\pm \tau)} & \ x>0,~-x<t<x \quad \text{Right Rindler}\\
-\frac{1}{\kappa} \ e^{-\kappa(\sigma\mp \tau)}& \ x<0,~x<t<-x\quad \text{Left Rindler}\end{array}\right.\nonumber
\end{equation}
and a similar relation for the Top/Down Milne. Note crucially that Diff $\!\oplus\!$ Weyl  invariance of the string action \eqref{worldsheet-action} implies  classical equivalence of the above worldsheet geometries.

Fig.~\ref{fig:Quadrants} shows Minkowski space. Here we are interested in a closed string whose worldsheet is a cylinder, and hence the Rindler/Milne maps only represent a part of the worldsheet. For this case, the four quadrants R, T, L and D  are related by $\mathbb{Z}_2\times \mathbb{Z}_2$ null orientifoldings:  $\tau\pm \sigma \to 2\pi -(\tau\pm\sigma)$. Explicitly, 
\begin{equation}\label{RTLD-mapping}
\begin{split}
R\to T:\ 
\tau+\sigma\to 2\pi-(\tau+\sigma),\qquad \tau-\sigma\to \tau-\sigma 
\\
T\to L:\ 
\tau+\sigma\to \tau+\sigma,\qquad \tau-\sigma\to 2\pi-(\tau-\sigma)  
\\
 L\to D:\ 
 \tau+\sigma\to 2\pi-(\tau+\sigma),\qquad \tau-\sigma\to \tau-\sigma 
\\
D\to R:\  
\tau+\sigma\to \tau+\sigma,\qquad \tau-\sigma\to 2\pi-(\tau-\sigma)
\end{split}
\end{equation}
Under both of the $\mathbb{Z}_2$'s, $(\tau,\sigma)\to (-\tau,2\pi-\sigma)$, which is the \textit{parity times time-reversal} (PT) at the level of worldsheet. 

In \eqref{x-t-sigma-tau} we have considered a uniformly accelerated worldsheet observer. One can consider a more general case of non-uniform acceleration for which the Weyl factor is  $e^{-2W(\tau,\sigma)}$, breaking both of the discrete symmetries. This gives rise to \textbf{Class IV} states. The case with constant $\partial_a W$ keep either (or both) of the $\mathbb{Z}_2$'s and yield \textbf{Class I} states (for $\partial_a W=0$), \textbf{Class II} states ($\partial_a W$ a constant timelike vector), \textbf{Class III} states ($\partial_a W$ a constant spacelike vector). At classical level  equations of motion for the pullback fields do not care about this $\mathbb{Z}_2\times \mathbb{Z}_2$ symmetry, and are same for Minkowski and Rindler. At quantum level, however, the local $\mathbb{Z}_2\times \mathbb{Z}_2$ which is a part of local Diff $\!\oplus\!$ Weyl symmetry should be carefully analyzed along with sandwich Virasoro condition \eqref{Sandwich-constriants-Tab}. That is what we do next. 

\begin{figure}[t]
\centering
\begin{tikzpicture}[scale=0.8]
\draw[gray, thin] (-2.5,0) -- (2.5,0);
\draw[->, gray] (2.5,0) -- (2.8,0);
\draw[blue] (2.8,0) node[above] (script) {$\sigma$};

\draw[blue] (1,0) node (script) {\large{\textbf{R}}};
\draw[blue] (-1,0) node (script) {\large{\textbf{L}}};

\draw[red] (0,1) node (script) {\large{\textbf{T}}};
\draw[red] (0,-1) node (script) {\large{\textbf{D}}};
\draw[black, thick] (-2,0) -- (0,2);
\draw[black, thick] (-2,0) -- (0,-2);
\draw[black, thick] (2,0) -- (0,2);
\draw[black, thick] (2,0) -- (0,-2);
\draw[black, thick] (-1,1) -- (1,-1);
\draw[black, thick] (-1,-1) -- (1,1);
\draw[gray, thin] (0,-2.5) -- (0,2.5);
\draw[->, gray] (0,2.5) -- (0, 2.8);
\draw[red] (0,2.8) node[right] (script) {$\tau$};
\end{tikzpicture}
    \caption{Penrose diagram of 2d Minkowski space is a 4-fold cover of R (or L or T or D) sectors. Each of the quadrants is conformal to Minkowski. L and R quadrants are two copies of what a Rindler observer have access to and T and D are two copies of a Milne space.    String worldsheet Rindler or Milne observers hence have access to Minkowski/$\mathbb{Z}_2\times \mathbb{Z}_2$. } 
    \label{fig:Quadrants}
\end{figure}
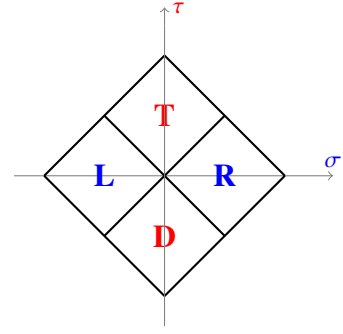

\textit{$\mathbb{Z}_2\times \mathbb{Z}_2$ and SVC.} Virasoro generators w.r.t the Minkowski observer $(\L^{\pm\text{M}}_{n})$  under worldsheet parity P and time-reversal T transform as, 
\be
\mathcal{L}_{n}^{\pm\text{M}} \xrightarrow{x\to -x} \mathcal{L}_{n}^{\mp\text{M}},\qquad 
\mathcal{L}_{n}^{\pm\text{M}} \xrightarrow{t\to -t} -\mathcal{L}_{-n}^{\mp\text{M}},
\ee
whereas the combinations
\be
J^{\text{M}}_n:= \mathcal{L}_{n}^{+\text{M}}+ \mathcal{L}_{ n}^{-\text{M}},\quad  P^{\text{M}}_n:= \mathcal{L}_{n}^{+\text{M}}- \mathcal{L}_{- n}^{-\text{M}}
\ee
are the only ones that do not mix under P and T,
\be
\begin{split}
\hspace*{-3mm} (J^{\text{M}}_n,P^{\text{M}}_n ) \xrightarrow{x\to -x} (J^{\text{M}}_n, -P^{\text{M}}_{-n} ),\ \  
(J^{\text{M}}_n,P^{\text{M}}_n ) \xrightarrow{t\to -t} (-J^{\text{M}}_{-n}, P^{\text{M}}_{n} ).
\end{split}\ee
Thus, $J^{\text{M}}_n$ and $P^{\text{M}}_n$ are respectively invariant under P and T. The constraints \eqref{SVC} are then explicitly equivalent to sandwich conditions for ${P}^{\text{M}}_{n}, {J}^{\text{M}}_{n}$, and \eqref{L0BL0-Sand--Const} yields 
\be \langle phys'|{P}^{\text{M}}_{0} |phys\rangle=0=\langle phys'|{J}^{\text{M}}_{0} |phys\rangle.
\ee 
Therefore, as discussed below \eqref{Class-I-II-III-IV-explicit}, \textbf{Class I} states are invariant under both P, T, \textbf{Class II} (or \textbf{III}) states 
that keep  T (or P), are  zero-eigenstates of  ${J}^{\text{M}}_{0}$ (or ${P}^{\text{M}}_{0}$) and \textbf{Class IV} states keep neither of P and T. See Table. \ref{Z2Z2-table} again for a summary.
 
One can use \eqref{RTLD-mapping} to map R, T, L and D Virasoro generators to each other, leading to the full set of identifications:
\begin{equation}\label{Ln-RTLD}
\mathcal{L}_{n}^{\pm\text{T}}=\mp\L^{\pm\text{R}}_{\mp n},\qquad
\mathcal{L}_{n}^{\pm\text{L}}=-\L^{\pm\text{R}}_{- n},\qquad 
\mathcal{L}_{n}^{\pm\text{D}}=\pm\L^{\pm\text{R}}_{\pm n},
\end{equation}
In a similar way, combinations \footnote{Note further that all of these combinations of generators have the same symmetry algebra:
\[\begin{split}
[{J}_n, {J}_m] &= (n-m){J}_{n+m}+\frac{c_1}{6} (n^3-n)\delta_{m+n,0},\\ [{J}_n,{P}_m] &= (n-m){P}_{n+m},\\ [{P}_n,{P}_m] &= (n-m){J}_{n+m}+\frac{c_2}{6} (n^3-n)\delta_{m+n,0},
\end{split}\]
which is just a rewriting of the conformal algebra with $c_1 = c -\bar{c}$ and $c_2 = c+\bar{c}$.} 
\begin{equation}\label{precarrolgen}
\begin{split}
{P}^{\text{R}}_{n}=\L^{+\text{R}}_{n}+\L^{-\text{R}}_{-n}=\L^{+\text{R}}_{n}-\L^{-\text{L}}_{n},\\ {J}^{\text{R}}_n=\L^{+\text{R}}_{n}-\L^{-\text{R}}_{-n}=\L^{+\text{R}}_{n}+\L^{-\text{L}}_{n}
\end{split}\end{equation}
do not mix under P and T and are invariant under either of P or T. In the second equalities above we have used \eqref{Ln-RTLD}. Moreover, using \eqref{Ln-RTLD} one may construct similar combinations for ${P}^{\text{T}}_{n}, {P}^{\text{L}}_{n}, {P}^{\text{D}}_{n}$ and similarly for their $J_n$ counterparts.  The above analyses suggest that to describe the same physics as seen by Minkowski observer with \textbf{Class I} Hilbert space (textbook string theory), a Rindler observer should use \textbf{Class I} and \textbf{III} Hilbert spaces and a Milne observer that of \textbf{Class I} and \textbf{II}. In what follows we establish this proposal.

\bec
\textbf{Beyond Class I states}
\label{sec:Unruh-Class-III}
\eec

Quantum mechanical theory of non-inertial (accelerated) worldsheets require the use of worldsheet Bogoliubov transformations. There is also an associated worldsheet Unruh effect: non-inertial worldsheet observer should use a physical Hilbert space different than a Minkowski observer \cite{Bagchi:2019cay, Bagchi:2020ats,Bagchi:2021ban} \footnote{For some older work along this direction, see \cite{deVega:1987um, deVega:1987veo} and references therein.}.  Bogoliubov transformations relating  a worldsheet Rindler observer with acceleration $\kappa$ and a Minkoswki one is given in terms of the unitary {\em squeezing operator} \cite{Bagchi:2019cay}
\begin{subequations}
\begin{align}\label{beta-alpha}
    \b_n(\kappa) &= e^{iG(\kappa)} \a_n  e^{-iG(\kappa)} , \ \ \tb_n (\kappa) = e^{iG(\kappa)} \ta_n  e^{-iG(\kappa)} 
    , \\
G(\kappa)&=2\kappa\left[\a_0 \cdot x-\ta_0\cdot \tilde{x}-\sum_{n=1}^\infty \frac{{i}}{n}  \left(\alpha_{-n}\tilde{\alpha}_{-n}-\alpha_{n}\tilde{\alpha}_{n}   \right)\right].\label{G-kappa}
\end{align}
\end{subequations}
The associated vacua are related as $|0(\kappa);k^\mu\rangle_{_\b} = e^{iG(\kappa)} |0; k^\mu\rangle_{_\a}$.
Minkowski and Rindler Virasoro generators are then related as $\L^{\pm{\mathcal{R}}}_n  = e^{iG(\kappa)} \L^{\pm\text{M}}_n e^{-iG(\kappa)}$ yielding
\begin{equation}\label{M-R-mapping}
\begin{split}
J^{\mathcal{R}}_n=e^{iG(\kappa)}\ J^{\text{M}}_n\ e^{-iG(\kappa)},\qquad P^{\mathcal{R}}_n=e^{iG(\kappa)}\ P^{\text{M}}_n\ e^{-iG(\kappa)}.
\end{split}
\end{equation} 
In particular, 
\be\label{R-M-mapp}
P^{\mathcal{R}}_0=P^{\text{M}}_0=N-\tilde{N},\quad J^{\mathcal{R}}_0= \cosh2\kappa\ J^{\text{M}}_0-\sinh2\kappa\ {\cal T}_0,
\ee
with $J^{\text{M}}_0=2\a_0^2 + N+\tilde{N}-2a$, $N=2\sum_{m>0} \a_{-m}\cdot\a_{m} \ ,  \tilde{N}=2\sum_{m>0} \ta_{-m}\cdot\ta_{m},$ and
\be
{\cal T}_{0}=2\sum_{m>0} \a_{-m}\cdot\ta_{-m}+ h.c.
\ee
where there a mixture of left and right Minkowski oscillators $\alpha_n, \tilde\alpha_n$, a feature induced from  $G(\kappa)$ \eqref{G-kappa}.

We now explore how a \textbf{Class I} state in the Minkowski frame $|r,r,k^\mu\rangle$ with $-k^2=4(r-a)$ is seen by in the Rindler observer. Recalling  that
$P^{\text{M}}_0|r,r,k^\mu\rangle=0=J^{\text{M}}_0 |r,r,k^\mu\rangle$, we can deduce,
\begin{subequations}\label{R-Class-I--III}
\begin{align}
P^{\mathcal{R}}_0 |r,r,k^\mu\rangle&=0,\\ 
J^{\mathcal{R}}_0 |r,r,k^\mu\rangle \sim & -\sinh2\kappa \sum_{\tilde{r}+\tilde{s}=2r} \Big[|\tilde{r},\tilde{r},k^\mu\rangle +|\tilde{s},\tilde{s},k^\mu\rangle\Big] \label{2}
\end{align}
\end{subequations}
{In \eqref{2}, we see the emergence of \textbf{Class III} states from \textbf{Class I} states, and the mass-shell condition of \textit{Rindler} \textbf{Class I} state becomes just a scaled version of the Minkowski cousin. Here, $\sim$ denotes equality up to some extra states that are in the complement Hilbert space of \textbf{Class I} states. Note that $J^{\mathcal{R}}_0 |r,r,k^\mu\rangle$ is a state that is orthogonal to all \textbf{Class I} states,  
thus the SVC are naturally satisfied even in the non-inertial setting  {\footnote{In \eqref{2}, we have dropped certain offshell states which in the sandwich conditions naturally go to zero. We believe these offshell remnants of type III states are the cue to something much deeper and their conjugates would show up in the corresponding Milne wedge leading to a stringy Hawking effect. We hope to report on this elsewhere.}},
\footnote{The $\beta$-vacuum $|0(\kappa);k^\mu\rangle_{_\b} = e^{iG(\kappa)} |0; k^\mu\rangle_{_\a}$ is a squeezed state \cite{Bagchi:2019cay} that satisfies SVC for $k^2=-m^2=4a$, i.e. for $m^2=-4a$ it is the sum of the $\alpha$-vacuum state and level-matched states of the form $|m,m,k^\mu\rangle$ that are in ${\cal H}_{\text{\tiny{C}}}$ complement to \textbf{Class I} Hilbert space.}}.

One can readily extend the above analysis for a Milne frame using \eqref{Ln-RTLD} and \eqref{precarrolgen}, and show that \textbf{Class I} states of a Minkowski observer yield \textbf{Class II} states of Milne observer. And for a generic non-inertial worldsheet, {with a mode number dependent $\kappa$,} one needs  \textbf{Class IV} states.

A similar description would also hold when \textbf{Class I} states in Rindler/Milne are observed from the Minkowski frame. We thus see that considering thermal aspects on non-inertial worldsheets, i.e. usual \textbf{Class I} string theory observed from an accelerated frame necessitates the inclusion of other classes of states that we had earlier discussed in a generic setting. It is important to emphasize that the SVC continues to hold in all frames for all states. This universality of SVC, irrespective of observers and states, is what we will call the \textit{``Worldsheet Equivalence Principle"}. It means in order that the same physics is observed by worldsheet observers from different frames,  these observers should use different types of states. It is also curious to note that in usual quantum field theory, the thermal observers typically have access to less states than their inertial counterparts, but with a thermal density matrix. For the case of  string theory, with an inherent non-locality, accelerated worldsheet observer requires more states.

\bec
\textbf{Concluding Remarks}\label{sec:conclusion}
\eec

We revisited the textbook analysis of quantum consistency of string worldsheet theory 
that requires imposing Sandwich Virasoro Conditions (SVC) \eqref{SVC}. Our careful analysis of the latter revealed that  \eqref{SVC} has  solutions beyond than the usually discussed ones \eqref{ra}. We established that solutions to \eqref{SVC} come in four classes, among which solutions to \eqref{ra} constitute \textbf{Class I} states and that the other solutions are required for having a consistent statement of equivalence for noninertial worldsheets at quantum level: The Hilbert space of a worldsheet uniformly accelerated observer should contain \textbf{Class I} and \textbf{Class III} states if she wants to describe the same physics as the usual worldsheet Minkowski observer who uses a Hilbert space consisting of only \textbf{Class I} states. This is a consequence of the fact that $G(\kappa)$ \eqref{G-kappa}, which maps Minkowski and Rindler operators,  mixes left and right oscillators. 

While 2d Rindler and Minkowski metrics are (locally) the same up to a Weyl factor, as depicted in Fig.~\ref{fig:Quadrants}, the Rindler covers only a quarter of Minkowski space. There is a $\mathbb{Z}_2\times \mathbb{Z}_2$ symmetry associated with worldsheet parity P and time-reversal T and even while considering the Weyl factor, Rindler and Minkowski worldsheets are \textit{distinguishable} through their behavior under this  $\mathbb{Z}_2\times \mathbb{Z}_2$. This was the key in our construction of solutions to SVC and understanding physics of non-inertial worldsheets. 

{A crucial remark is that \textbf{Class II, III} or \textbf{IV} states are not level-matched or on-shell in the  usual string theory terminology. Our explicit construction exhibits that these states are necessitated  with horizons on the worldsheet that appear through ``null-orientifolds'' \eqref{RTLD-mapping}. An important take home message is, the SVC (and not just Virasoro highest-weight representations) are absolutely indispensable in understanding the quantum nature of such worldsheet horizons.
It has been argued that worldsheet Rindler observers are ultimately apt to study strings probing black hole horizons in the target spacetime \cite{Bagchi:2020ats, Bagchi:2021ban, Bagchi:2023cfp, Bagchi:2024rje}, or when one heats up a gas of strings. Thus, our analysis and results here, besides adding a couple of pages to string theory textbooks, is crucial to study strings probing black holes 
as well as the still mysterious Hagedorn phase transition \cite{Hagedorn:1965st, Atick:1988si}.  Since the physics of both cases seemingly rely on \textit{Carrollian} symmetries \cite{NDS, LBLL} on the worldsheet 
\cite{Bagchi:2013bga, Bagchi:2015nca}, it would be very important to connect this construction to such symmetries at the quantum level.}

\textbf{Acknowledgements.} We would like to thank Daniel Grumiller, Ashoke Sen and Joan Sim\'on for discussions and comments on the manuscript.

ABag is partially supported by a Swarnajayanti Fellowship from the Science and Engineering Research Board (SERB) under grant SB/SJF/2019-20/08 and also by SERB grant CRG/2022/006165. ABan is supported in part by an OPERA grant and a seed grant NFSG/PIL/2023/P3816 from BITS-Pilani  and an early career research grant ANRF/ECRG/2024/002604/PMS from ANRF India. He also acknowledges financial support from the Asia Pacific Center for Theoretical Physics (APCTP) via an Associate Fellowship. MMShJ is supported in part by Iran National Science Foundation (INSF) Grant No. 4026712.
He also gratefully acknowledges the support from the Abdus Salam ICTP, through Senior Associates Programme (2023-2028) and ICTP HECAP section.

\bibliographystyle{JHEP}
\bibliography{reference}

\end{document}